\documentclass[twocolumn,a4paper]{revtex4-1}
\usepackage{hyperref}
\usepackage{graphicx}
\usepackage{amsmath}
\usepackage{xcolor}
\usepackage{float}
\usepackage{amsfonts}
\usepackage{epsfig}
\usepackage{graphics}
\usepackage{color}
\usepackage{csquotes}

\newcommand{\beq}{\begin{equation}}
\newcommand{\eeq}{\end{equation}}
\newcommand{\bqa}{\begin{eqnarray}}
\newcommand{\eqa}{\end{eqnarray}}

\begin{document}
\title{Heavy quarkonium states with baryonic chemical potential}
\author{Siddhartha Solanki$^{a}$}
\author{Manohar Lal$^{a}$} 
\author{Vineet Kumar Agotiya$^{a}$}
\email{agotiya81@gmail.com}
\affiliation{$^a$Department of Physics, Central University of Jharkhand, Ranchi, India, 835-222}

\begin{abstract}
In this work, we have studied the dissociation behavior of 1S and 2S states of quarkonium using quasi-particle approach where the Debye mass depends on baryonic chemical potential. The binding energies of the quarkonium states has been obtained by using quasi-particle Debye mass which further depends on temperature and baryonic chemical potential ($\mu_{b}$). The effect of $\mu_{b}$ on the binding energies and the dissociation temperature have been also studied. The binding energy and dissociation temperature of heavy quarkonia decreases as $\mu_{b}$  increases. The effect of $\mu_{b}$ on the mass spectra of quarkonium states has been studied well.\\
\end{abstract}

\maketitle

\section{Introduction}
The physics of the elementary particles played a key role to  understand the quantum-chromodynamics (QCD) which is considered as the theory of strong interaction between the quarks and gluons.The QCD theory helps to understand the different phases/events occurring at different temperatures and baryon densities. For example, at small or vanishing temperature gluons and quarks are limited by the Browny forces while at higher temperature, asymptotic freedom advocate a somewhat distinct QCD medium which consists of weakly coupled deconfined gluons and quarks known as the quark-gluon plasma (QGP). Suppression of the $J/\psi$ (ground state of charmonium) meson was move back as a feasible unmistakable sign of the beginning of deconfinement. Matsui and Satz ~\cite{M.C.Abreu,T.Matsui}argued that charmonium states, generated before the formation of a thermalized QGP, would tend to melt in their way through the deconfined medium, because the coulombic potential is screened by the large number of color charges, thereby producing an anomalous drop in the  $J/\psi$ yields. The pairs develops into the physical resonance during formation time and passes through the plasma and hadronic matter before they leave the interacting system to decay into a dilepton to detected. Even before the resonance occurs, $J/\psi$ meson may be engaged by the combination of protons and neutrons (nucleons) pouring past it~\cite{C.Gerschel}. By that time the delocalization is formed, color screening in the QGP may be adequate to inhibit a binding of charmonium~\cite{T.Matsui}, an dynamic gluon~\cite{X.M.Xu}, and a co-moving hadron could separate the resonance. The  study of quarkonia pair Q$\bar{Q}$ at finite values of temperature is of immense/utmost important for studying QGP formation in HIC's. Many efforts have been made to determine the dissociation temperature ($T_D$) of quarkonium state in the deconfined medium, using either lattice calculations of $Q\bar{Q}$ spectral function or non-relativistic calculations based upon some effective screened potential.\par
Lattice studies are directly based on QCD and should provide, in principle, a definitive answer to the problem. However in lattice studies, the spectral function must be pull out using limited sets of data from the Euclidean correlators, are directly based on the open framework. This along with the intrinsic technical issues of open framework calculations, limit the solidness of the results earned so far, and also their scope, which is basically restricted for the mass of the ground states in one and all quarkonium channel. The potentials model  provides a direct information the properties of quarkonia (i.e., charmonium and bottomonium) at finite values of temperature, by means of which we can calculate those quantities which are beyond the scope of lattice QCD approaches. Umeda and Alberico~\cite{T.Umeda, W.M.Alberico(2008)} have shown that the lattice computations of mesons correlators at finite values of temperature hold a constant contribution because of the existence of zero modes in the spectral functions.\par
The dissociation process of heavy  quarkonia in hot QCD medium has its importance in HIC's due to the fact that it provides sufficient information about the creation of QGP~\cite{Leitch}. In the last couples of year, the understanding about dissociation of quarkonium without any restricted medium has gone through several studies~\cite{Laine:2008cf,BKP:2000,BKP:2001,BKP:2002,BKP:2004}. As we know, in the quarkonium state the quarks-antiquarks are bound together by almost static (off-shell) gluons, therefore, the issue of their dissociation boils down to how the gluon self-energy behaves at high temperatures. It has been noticed that the gluon self-energy has both real and imaginary parts~\cite{laine}. 
\begin{figure*}
    \vspace{2mm}   
    \includegraphics[height=6cm,width=7cm]{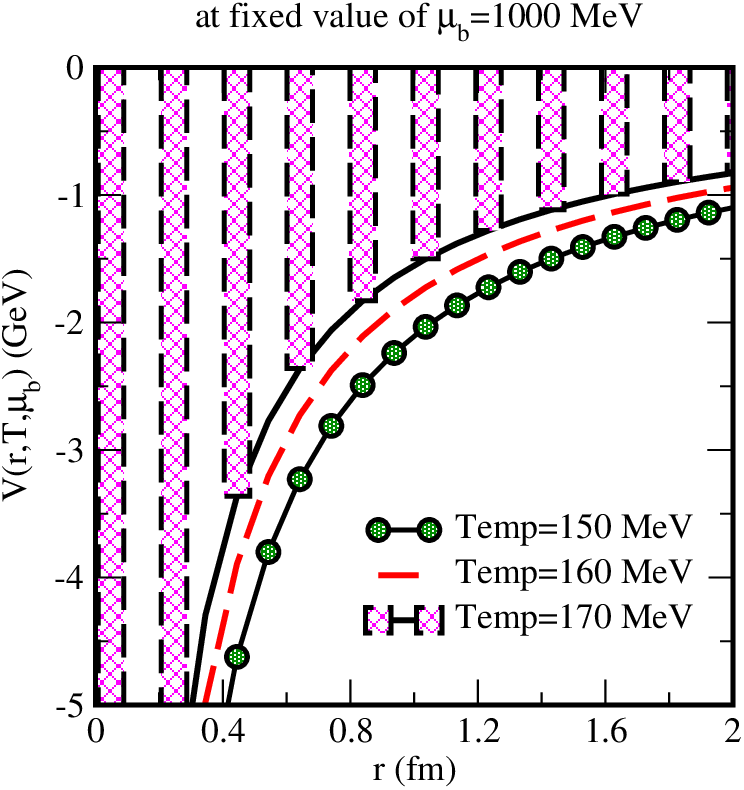}
    \hspace{12mm}
    \includegraphics[height=6cm,width=7cm]{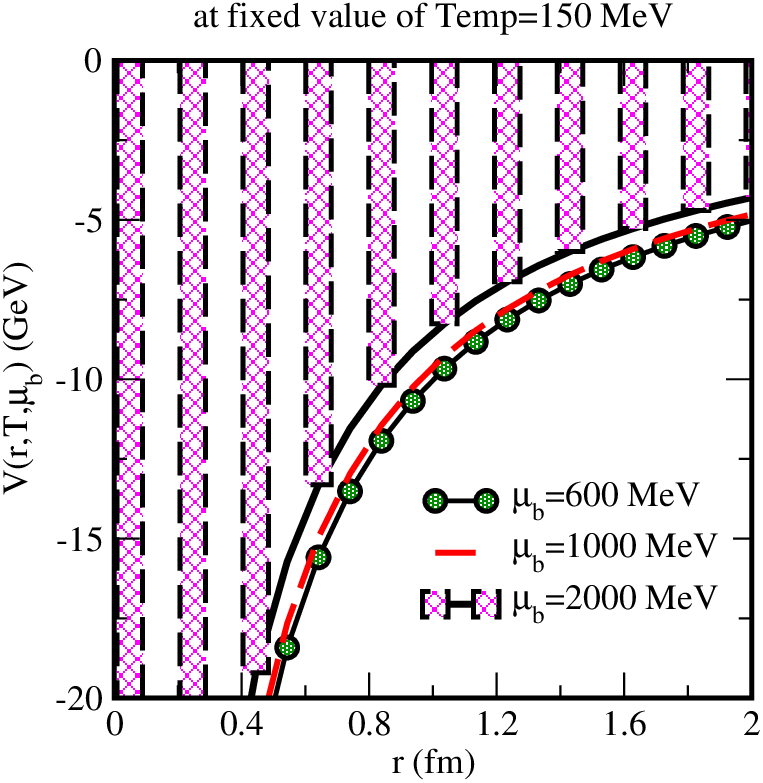} 
    \vspace{2mm}
\caption{The variation of Cornell potential in a hot and dense QCD medium at fixed value of $\mu_{b}$ with different values of temperatures (left panel) and fixed value of temperature with different values of $\mu_{b}$ (right panel)}.
\label{Figure.1}
 \vspace{2mm}   
\end{figure*}
The real part of gluon self-energy leads to  Debye screening, and the imaginary part of gluon self-energy leads to landau damping and give to the thermal width for the study of quarkonium properties. At higher values of temperature, QCD deconfined phase of matter undergoes static color-screenings~\cite{E.V.Shuryak,GPY}. In that case, it is expected that the screening will conduct the dissociation of the states of quarkonia. The potential model descriptions are also applicable for the study of the quarkonium properties at finite values of temperature and baryonic chemical potential.\\
Note that, the formation of ground states of quarkonium mesons in the hadronic reactions occurs in lump through the production of $\psi^{\prime}$ and $\Upsilon^{\prime}$ states of quarkonia and their decay into the specific ground state. Since the lifetime of different states of quarkonium is wide-ranging than the classic duration of life of the medium fabricated in nucleus-nucleus collisions; so their decays occurs completely different than produced medium~\cite{lain,he1}. The manufactured medium can be examine not only by the J/$\psi$ and $\Upsilon$ but also by the $\psi^{\prime}$ and $\Upsilon^{\prime}$. The representation of potential in this circumstances could be helpful in forecasting the binding energies of various states of quarkonia state by build up and solving proper Schrodinger equation in the hot QCD medium. The first step towards this is to model an appropriate medium dependent inter-quark interaction potential at finite temperature and baryonic chemical potential. Thereafter, the dissociation of excited states of quarkonium has been studied.\\ 
This manuscript is organized as follows: Study of potential using finite $\mu_{b}$ has been discussed in section (II). Whereas in section (III), Quasi-Particle Debye mass with $\mu_{b}$ in the hot QCD medium have been discussed. In section (IV), The effect of the $\mu_{b}$ on the binding energy and dissociation temperature of quarkonium states have been studied. The mass spectra of the quarkonium states have been calculated in section (V). In section (VI) we discussed the results of the present work and finally, we have concluded our work in Section (VII).
   \begin{figure*}
   \vspace{2mm}   
    \includegraphics[height=6cm,width=7cm]{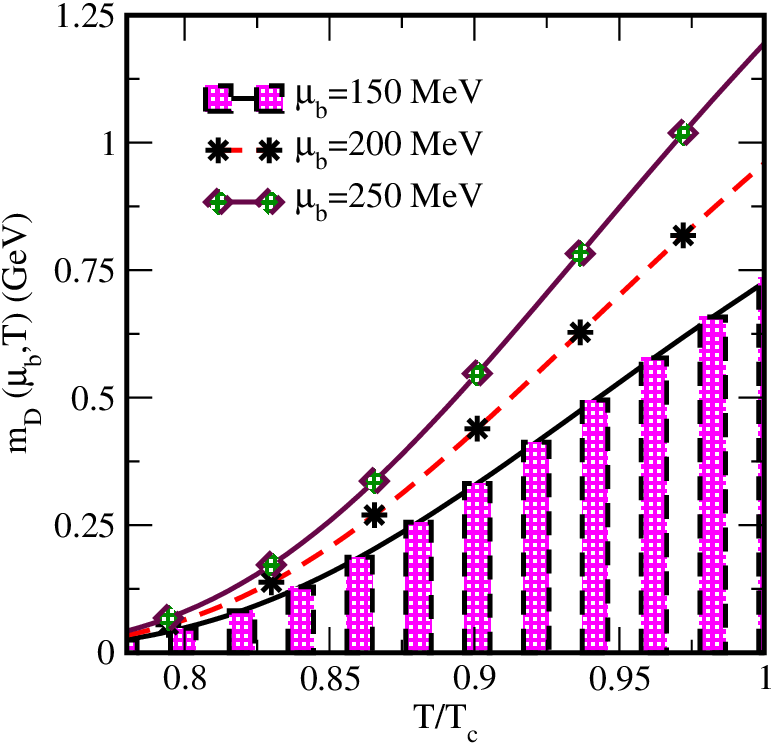}
    \hspace{12mm}
    \includegraphics[height=6cm,width=7cm]{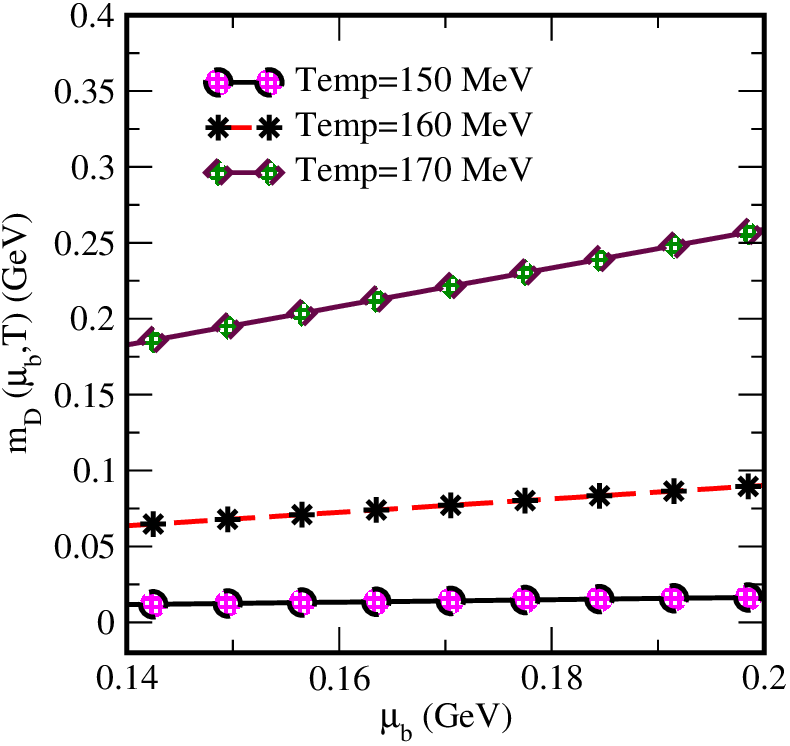}
    \vspace{2mm}
\caption{The variation of QP Debye mass with temperature at different values of $\mu_b$ (left panel) and with $\mu_b$ at different values of temperature (right panel)}.
\label{Figure.2}
 \vspace{2mm}   
\end{figure*}

\section{Modified form of Cornell potential with baryonic chemical potential ($\mu_{b}$)}
Proper understanding about the properties of quarkonium spectra requires interacting potential at finite values of temperature obtained directly from QCD, like the Cornell potential at zero value of temperature has been obtained from potential non-relativistic quantum chromodynamics (pNRQCD) along with the matching coefficient of zeroth-order. Such inferences at finite values of temperature for weakly coupled plasma have been come up in the literature recently~\cite{Brambilla05,Brambilla08} but they are, however, very sensitive to temperature soft as well as hard scales, $T$, $g^2 T$, $gT$, respectively. Due to these difficulty arises in the effective field theories (EFT) at finite temperature, the lattice based potentials become another choice to study the quarkonia spectra.\par 
However, neither the internal energy nor the free energy in the potential can be used directly. The potential model studies as well as the lattice QCD approach infer us that the interaction of quark antiquark potential plays a key role to understand the behavior of quark antiquark bound state in the hot QCD/QGP medium. The potential which employed is commonly screening coulomb (Yukawa form)~\cite{Brambilla05,L.Kluberg}. In  case of finite values of temperature, we engage the ansatz, the medium modification enter in the Fourier transform (FT) of heavy quark potential V(k) as~\cite{V.Agotiya:2009}.

\begin{equation}
\label{eq1}
\tilde{V}(k) \varepsilon(k)=V(k)
\end{equation}
Where, dielectric permitivity ($\varepsilon(k)$) is obtain by the static limit of longitudinal part of the gluon self energy~\cite{R.A.Schneider,H.A.Weldon}.
\begin{equation}
\label{eq2}
\varepsilon(k)\equiv\left(1+\frac{m^2_D(r,T,\mu_b)}{k^2}\right)
\end{equation}
V(k) is the FT of the Cornell potential, which is given below:
\begin{equation}
\label{eq3}
V(k)+\frac{4\sigma}{\sqrt{2\pi}k^4}=-\sqrt{\frac{2}{\pi}}\frac{\alpha}{k^2}
\end{equation}
Putting the values of  Eq.(\ref{eq2}) and Eq.(\ref{eq3}) in the Eq.(\ref{eq1}), and solving using inverse FT, we get the medium modified potential depending upon distance (r)~\cite{V.Chandra:2007,R.A.Schneider,A.Ranjan}.
\begin{multline}
\label{eq4}
V(r,T,\mu_{b})=\left(\frac{2\sigma}{m^2_D(T,\mu_{b})}-\alpha\right)\frac{exp(-m_D(T,\mu_{b})r)}{r}\\-\frac{2\sigma}{m^2_D(T,\mu_{b})r}+\frac{2\sigma}{m_D(T,\mu_{b})}-\alpha m_D(T,\mu_{b})
\end{multline}
Where, $\alpha$ is the coupling constant and $\sigma$=0.184 $GeV^{2}$ is the string coefficient.
\begin{figure*}
    \vspace{2mm}   
    \includegraphics[height=6cm,width=7cm]{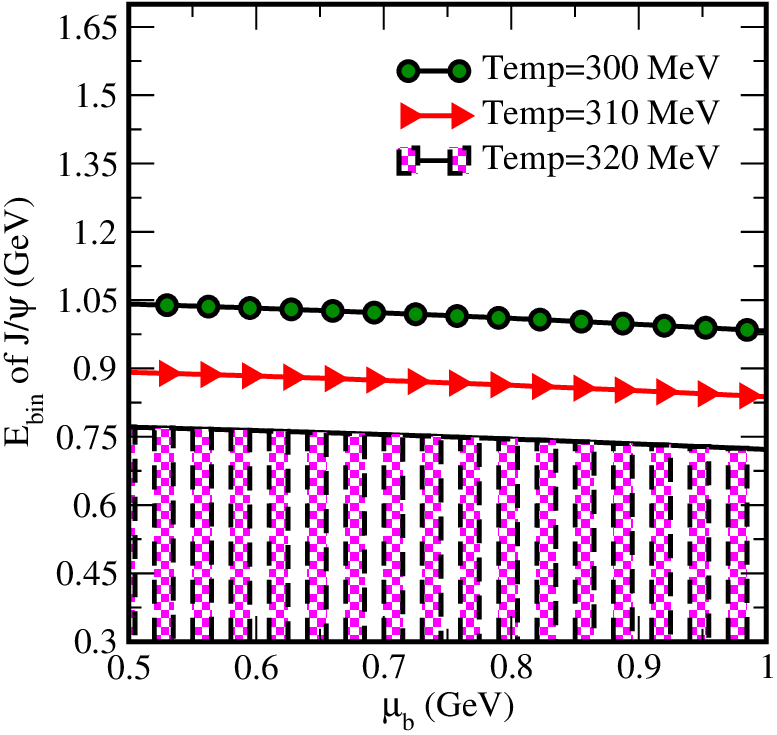}
    \hspace{12mm}
    \includegraphics[height=6cm,width=7cm]{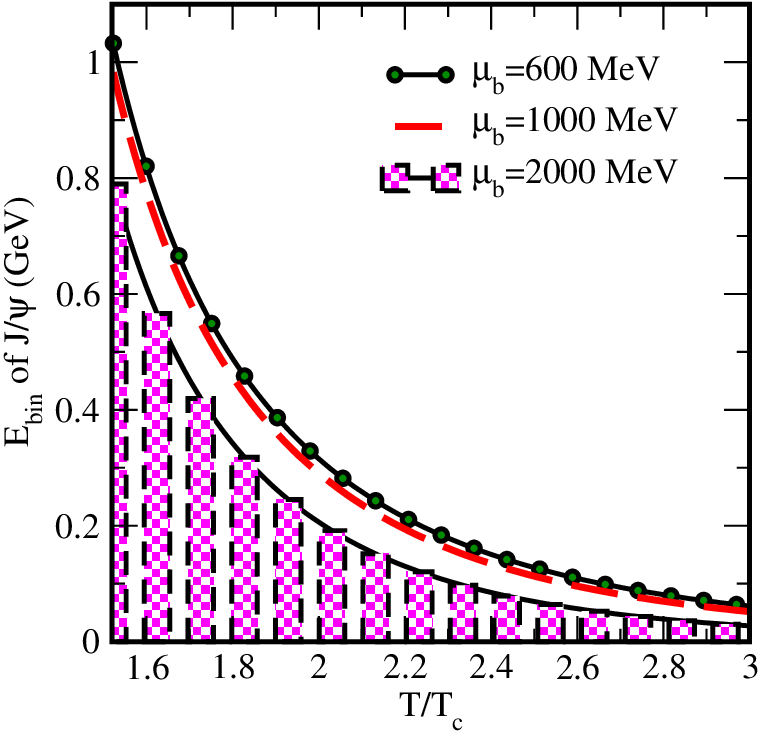}
    \vspace{2cm}
\caption{The variation of $E_{bin}$ of $J/\psi$ with $\mu_{b}$ at different values of temperatures (left panel) and with the temperature at different values of $\mu_{b}$ (right panel).}
\label{Figure.3}
\vspace{3cm} 
\end{figure*}
\begin{figure*}
    \vspace{2mm}   
    \includegraphics[height=6cm,width=7cm]{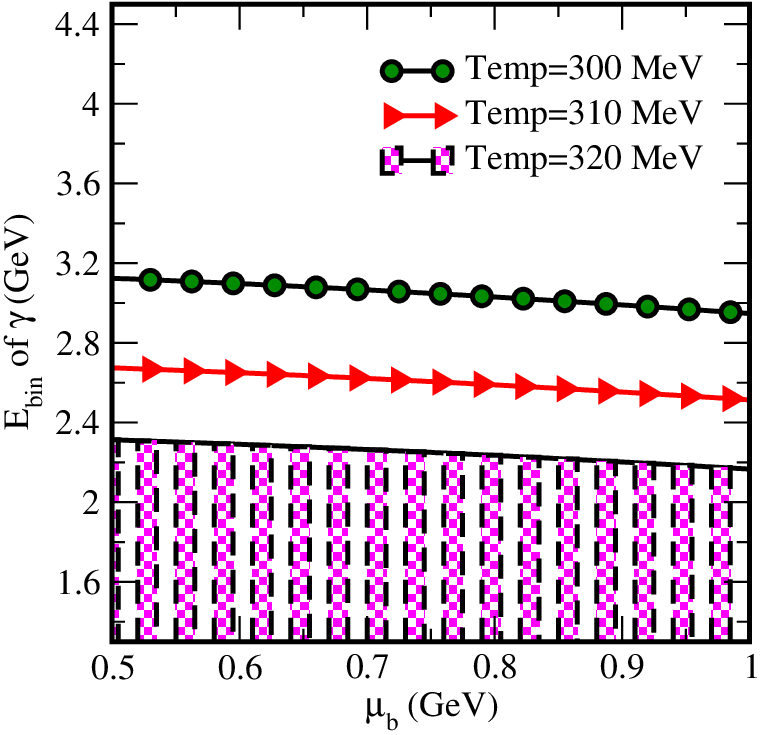}
    \hspace{12mm}
    \includegraphics[height=6cm,width=7cm]{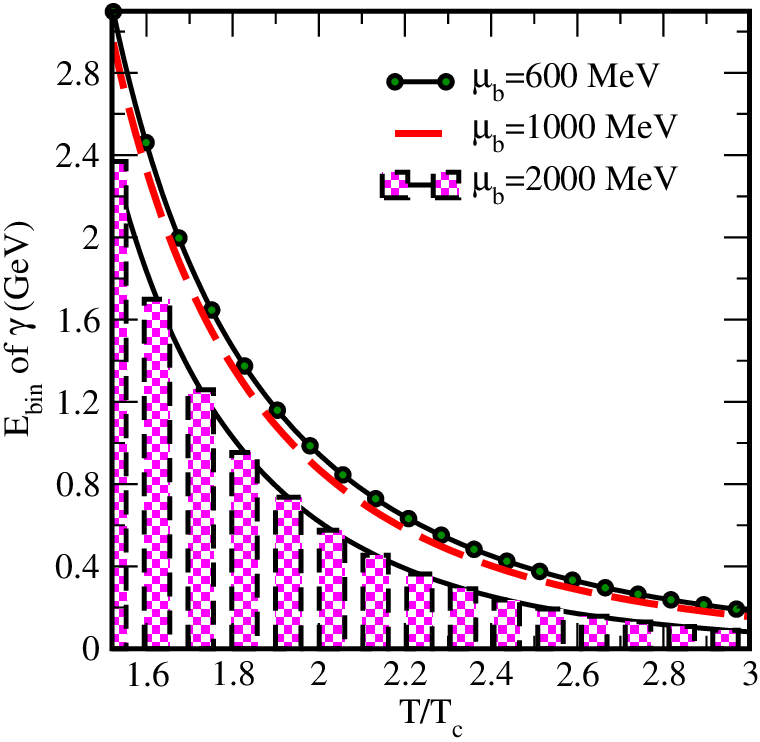}
    \vspace{2cm}
\caption{The variation of $E_{bin}$ of $\Upsilon$ with $\mu_{b}$ at different values of temperature (left panel) and with the temperature at different values of $\mu_{b}$ (right panel).}
\label{Figure.4}
\vspace{3cm} 
\end{figure*}

\section{Study of Quasi-Particle (QP) Debye mass with $\mu_{b}$ in the hot QCD medium}
The Leading order Debye mass (LO) has perturbative nature in QCD coupling at maximum range of temperature is known for a long time~\cite{A.Rebhan}. The gauge independent non-perturbative Debye mass in the QCD defined in~\cite{E.Braaten}. The calculation of Debye mass is mainly for the two Polykov loops by Braaten and Nicto at high temperature~\cite{Y.Burnier}. The basic definition of the Debye mass itself creates a difficulty because of the nature of gauge variant electric correlators in~\cite{K.Kajantie}. To get control of this problem many proposals have been purposed so far~\cite{K.Kajantie, Anbazavov, S.Nadkarni}. All the interaction between the quasi particle because of the quasi parton, a number of endeavor have been made so far such as, mass of effective model~\cite{V.Goloviznin,A.Peshier}, effective mass with Polyakov loop~\cite{M.D.Elia}, model based on PNJL and NJL~\cite{A.Dumitru}, effective fugacity model~\cite{V.Chandra:2007,V.Chandra:2009}. 
\begin{figure*}
    \vspace{2mm}   
    \includegraphics[height=6cm,width=7cm]{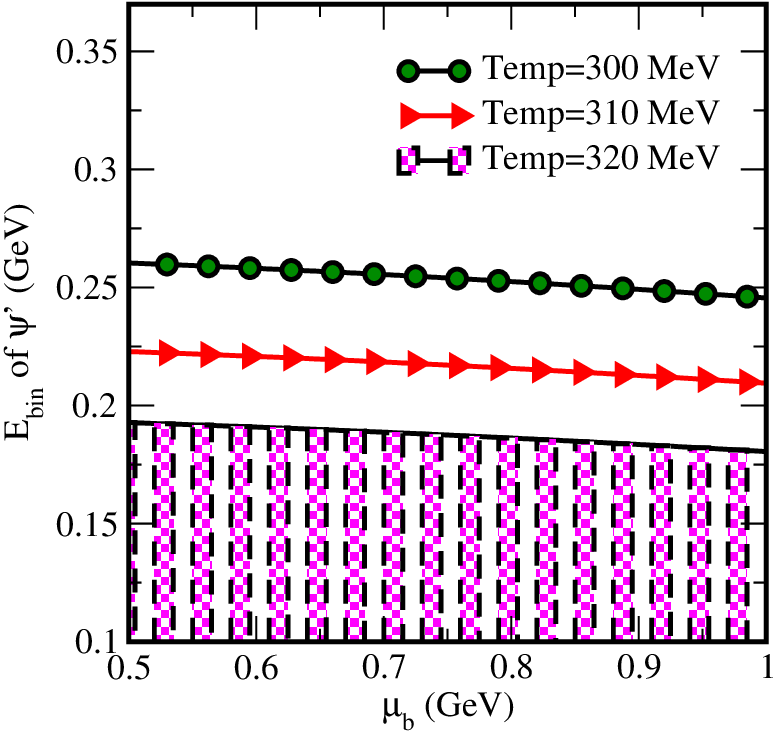}
    \hspace{12mm}
    \includegraphics[height=6cm,width=7cm]{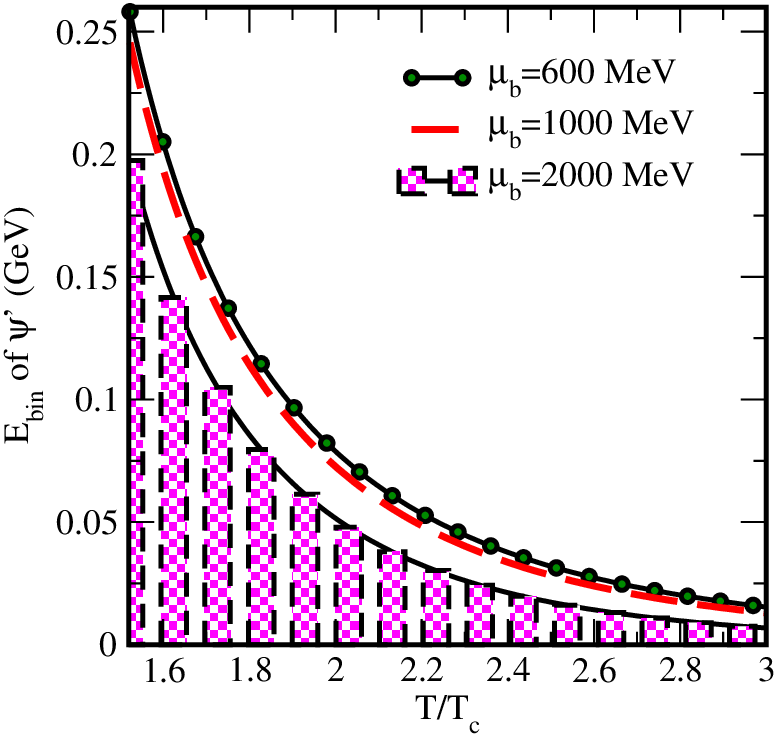}
    \vspace{2cm}
\caption {The variation of $E_{bin}$ of $\psi^{\prime}$ with $\mu_{b}$ at different values of temperature (left panel) and with the temperature at different values of $\mu_{b}$ (right panel)}
\label{Figure.5}
\vspace{3cm} 
\end{figure*}
\begin{figure*}
    \vspace{2mm}   
    \includegraphics[height=6cm,width=7cm]{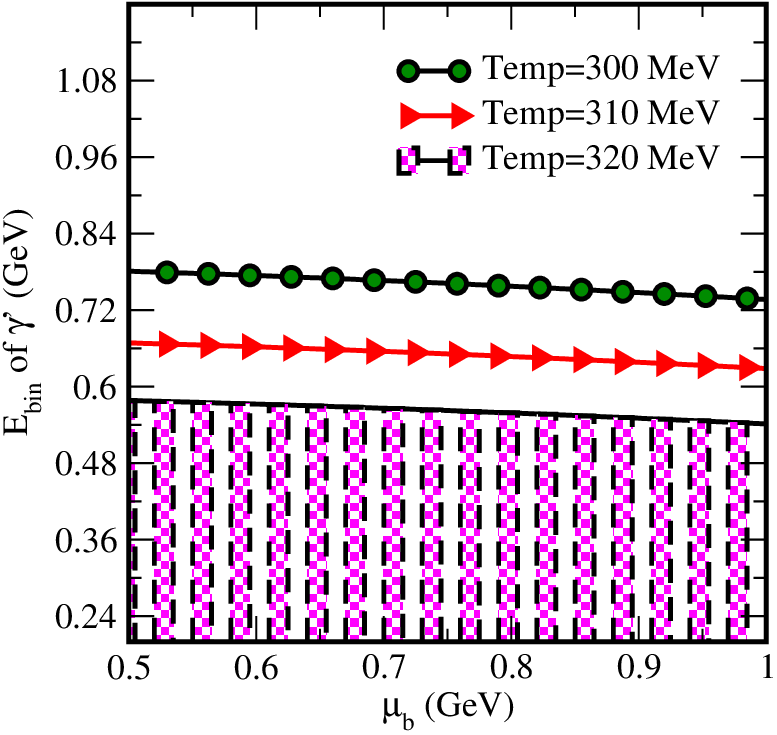}
    \hspace{12mm}
    \includegraphics[height=6cm,width=7cm]{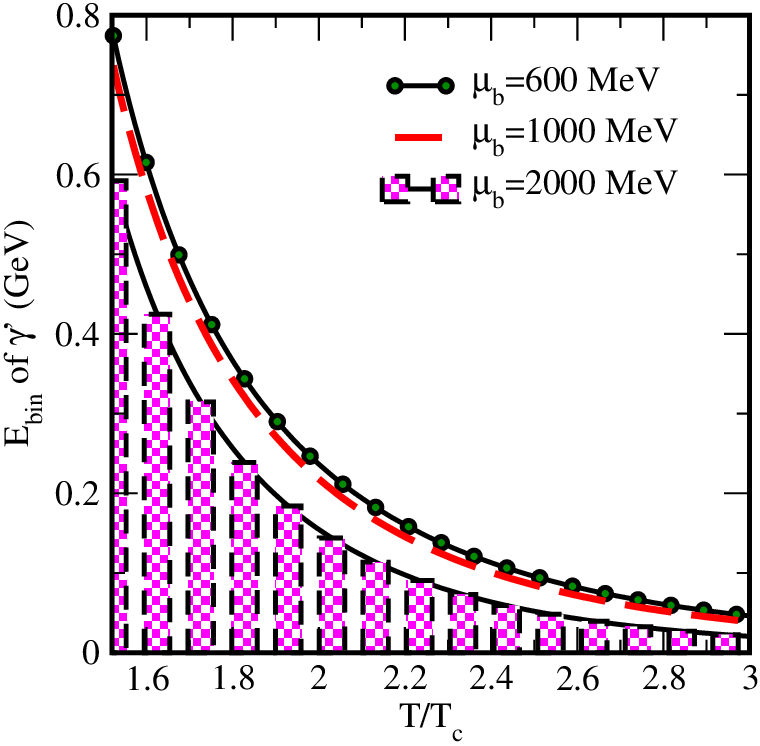}
    \vspace{2cm}
\caption{The variation of $E_{bin}$ of $\Upsilon^{\prime}$ with $\mu_{b}$ at various values of temperature (left panel) and with the temperature at various values of $\mu_{b}$ (right panel)}.
\label{Figure.6}
 \vspace{3cm}  
\end{figure*}
Quasi particle model is important key to define the non-ideal behavior of the QGP and the masses arises because of the neighbouring matter around the parton and this quasi parton obtained the identical quantum number as the real particle i.e gluons and quarks~\cite{P.K.Srivastava}. Here, we have used QP equation of states (EoS) as one can find in~\cite{M.Cheng}. The QP Debye mass $({m_D})$ for full QCD case is given as:
\begin{eqnarray}
\label{eq5}
m^2_D\left(T\right) &=& g^2(T) T^2 \bigg[
\bigg(\frac{N_c}{3}\times\frac{6 PolyLog[2,z_g]}{\pi^2}\bigg)\nonumber\\&&
+{\bigg(\frac{\hat{N_f}}{6}\times\frac{-12 PolyLog[2,-z_q]}{\pi^2}\bigg)\bigg]}
\end{eqnarray}
and the value of $\hat{N_f}$ is, as follow:
\begin{eqnarray}
\label{eq6}
\hat{N_f} &=& \bigg(N_f +\frac{3}{\pi^2}\sum\frac{\mu_{q}^2}{T^2}\bigg)
\end{eqnarray}
and $\mu_{q}$=$\frac{\mu_b}{3}$ as found in reference~\cite{U.Kakade}, where quark-chemical potential $(\mu_{q})$. Here, $g(T)$ is the QCD running coupling constant, $N_c$=$3$ ($SU(3)$) and $N_f$ is the number of flavor, the function $PolyLog[2,z]$ having form, $PolyLog[2,z]$=$\sum_{p=1}^{\infty} \frac{z^p}{p^2}$ and $z_g$ and $z_q$ is the quasi gluon effective fugacity and quasi quark effective fugacity. The isotropic nature of this distribution functions. The temperature dependence $z_g$ and $z_q$ fits well to the form given below:
\begin{figure*}
    \vspace{2mm}   
    \includegraphics[height=6cm,width=7cm]{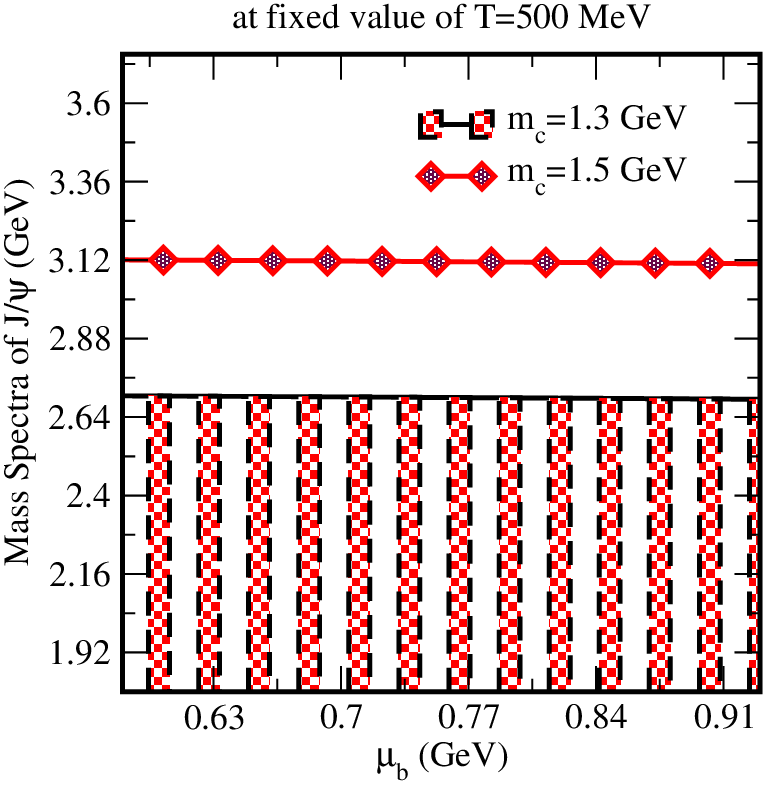}
    \hspace{12mm}
    \includegraphics[height=6cm,width=7cm]{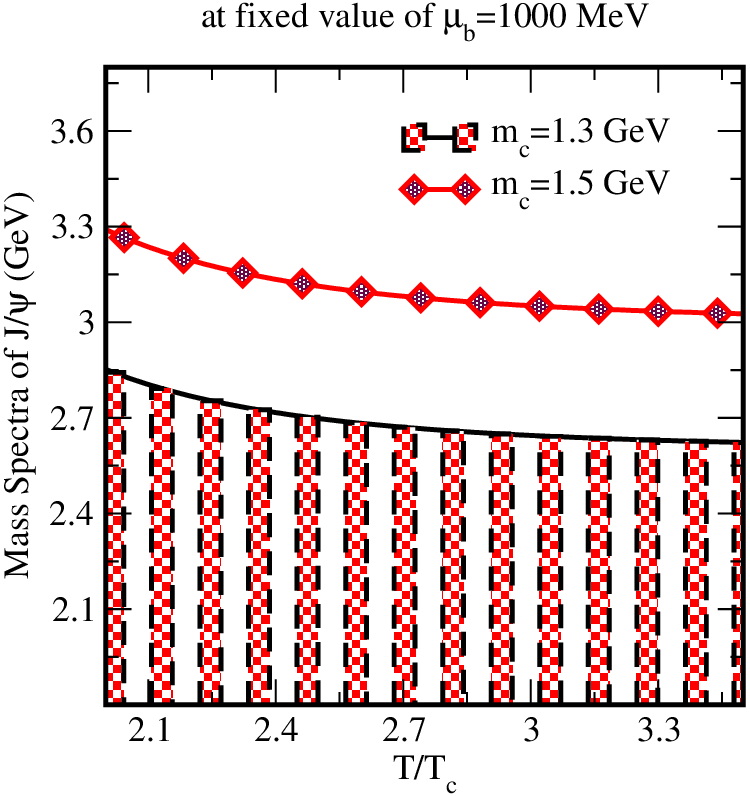}
    \vspace{2cm}
\caption{Dependency of mass of charmonium with $\mu_{b}$ (in left panel) and with $T/T_c$ (in right panel) for ground state of charmonium ($J/\psi$).}
\label{Figure.7}
 \vspace{3cm}  
\end{figure*}
\begin{figure*}
    \vspace{2mm}   
    \includegraphics[height=6cm,width=7cm]{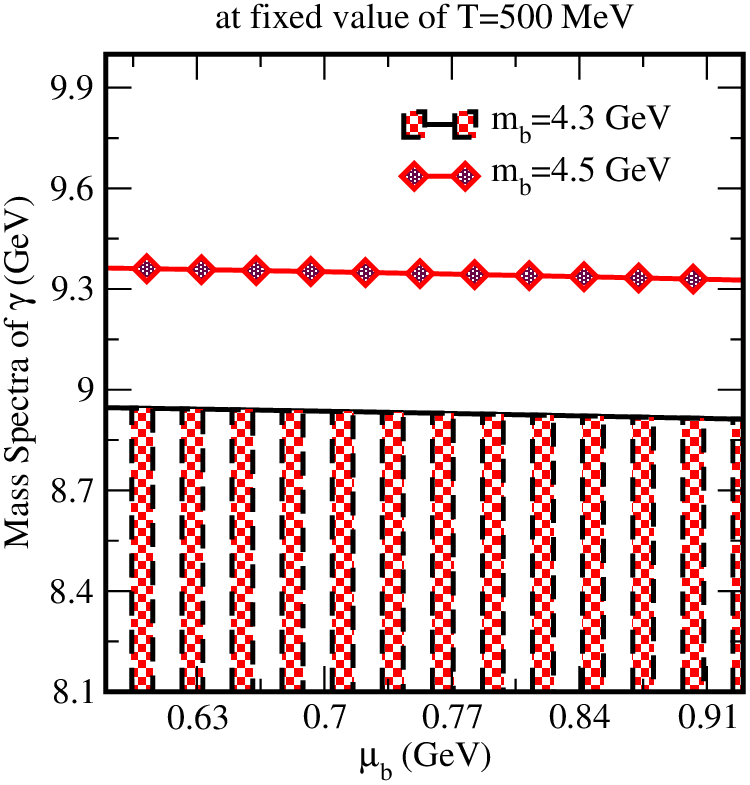}
    \hspace{12mm}
    \includegraphics[height=6cm,width=7cm]{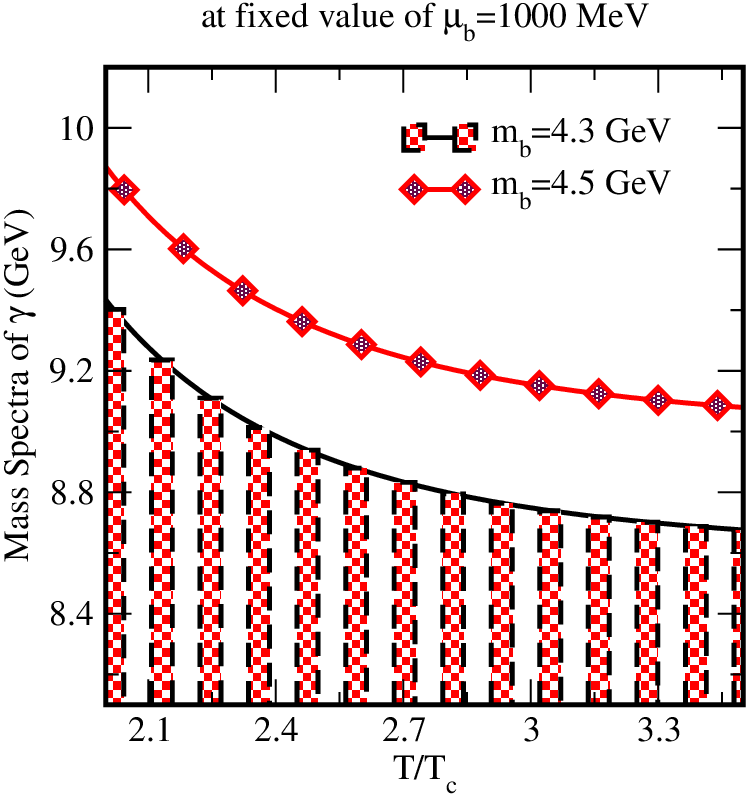}
    \vspace{2cm}
\caption{Dependency of mass of bottomonium with $\mu_{b}$ (in left panel) and with $T/T_c$ (in right panel) for ground state of bottomonium ($\Upsilon$).}
\label{Figure.8}
 \vspace{3cm}  
\end{figure*}
\begin{equation}
\label{eq7}
z_{g,q}= j_{q,g}\exp\bigg(-\frac{k_{g,q}}{y^2}-\frac{l_{g,q}}{y^4}-\frac{m_{g,q}}{y^6}\bigg).
\end{equation}
(Here y=$T/T_c$ and j, k, l and m are fitting parameters), for equation of state in the QP description~\cite{V.Chandra:2007,V.Chandra:2009} respectively. After introducing the value of $\hat{N_f}$ in the Eq.(\ref{eq5}), then the full declaration of the expression of QP Debye mass in terms of temperature and $\mu_{b}$~\cite{Solanki,Solanki2022} is:
\begin{equation}
\label{eq8}
m^2_D\left(T,\mu_{b} \right)=T^2\left\{\bigg[\frac{N_c}{3} Q^2_g\bigg]+\bigg[\frac{N_f}{6}+\frac{1}{2\pi^2}\bigg(\frac{\mu_{b}^2}{9 T^2}\bigg)\bigg] Q^2_q\right\}
\end{equation}
Where, the expression of the values of effective charges $Q_g$ and $Q_q$ is:
\begin{eqnarray}
\label{eq9}
 Q^2_g&=&g^2 (T) \frac{6 PolyLog[2,z_g]}{\pi^2}\nonumber\\
 Q^2_q&=&g^2 (T)  \frac{-12 PolyLog[2,-z_q]}{\pi^2}
\end{eqnarray}
In analysis, we have used the QP Debye mass, $(m_D^{QP})$ depending upon the temperature and $\mu_{b}$ for the full QCD case.

\section{Binding energy ($E_{bin}$) and Dissociation temperature of quarkonium state}
In this section, we have calculated the binding energy ($E_{bin}$) and dissociation temperature ($T_{D}$) of the ground and excited states of quarkonia after using the values of $\mu_{b}$. To reach this end we solve the Schrodinger equation for the complete understanding about the quarkonia in the hot QGP medium. The $E_{bin}$ of charmonium and bottomonium state at T=0 is defined by the difference of energy between the $m_Q$ (mass of quarkonia) and the bottom/open charm threshold. But distance between the continuum threshold and the peak position is defined the $E_{bin}$ at finite value of temperature~\cite{AgnesMocsy}. In our case, it is defined as the ionization potential because of the similarity of the above potential with the H-atom problem~\cite{T.Matsui}. So the time independent spherical Schrodinger equation gives the energy eigen values for the J/$\psi$ and $\Upsilon$ (ground states) and the $\psi^{\prime}$ and $\Upsilon^{\prime}$ (first excited states) charmonium and bottomonium spectra. By solving the Schrodinger equation we obtained the energy eigen values of the 1S and 2S states as below:\begin{equation}
\label{eq10}
 E_n = -\frac{1}{n^2}\frac{m_Q\sigma^2}{m^4_D}
\end{equation}
Where, the mass of the heavy quark is $m_Q$. The above expression of eigen values is called as the Ionization energy of the $n^{th}$ bound states. The effect of chemical potential enters through the Debye mass. It was observed that the $E_{bin}$ decreases with increasing the temperature and $\mu_{b}$ which is shown in the figures, \ref{Figure.3}, \ref{Figure.4}, \ref{Figure.5} and \ref{Figure.6}. When binding energy of charmonium and bottomonium state at particular values of temperature becomes smaller or equal to the value of mean thermal energy, the state which said to be dissociated. It was calculated by using the following expression $E_{bin}$=$T_{D}$ (for upper bound of quarkonium dissociation) and $E_{bin}$=$3T_{D}$ (for lower bound of quarkonium dissociation) i.e.,
\begin{eqnarray}
\label{eq11}
 \frac{1}{n^2} \frac{m_Q\sigma^2}{m^4_D}&=&3T_{D}\nonumber\\
 \frac{1}{n^2} \frac{m_Q\sigma^2}{m^4_D}&=&T_{D}
\end{eqnarray}
The dissociation temperature for the states of charmonium and bottomonium have been also discussed in~\cite{P.Sandin,prd2016_vin,prd2018_vin,SiddECS}. Here we have calculated lower and upper bound of dissociation pattern for the ground and excited state of charmonium and bottomonium using $\mu_{b}$ in table I and II respectively.
\begin{figure*}
    \vspace{2mm}   
    \includegraphics[height=6cm,width=7cm]{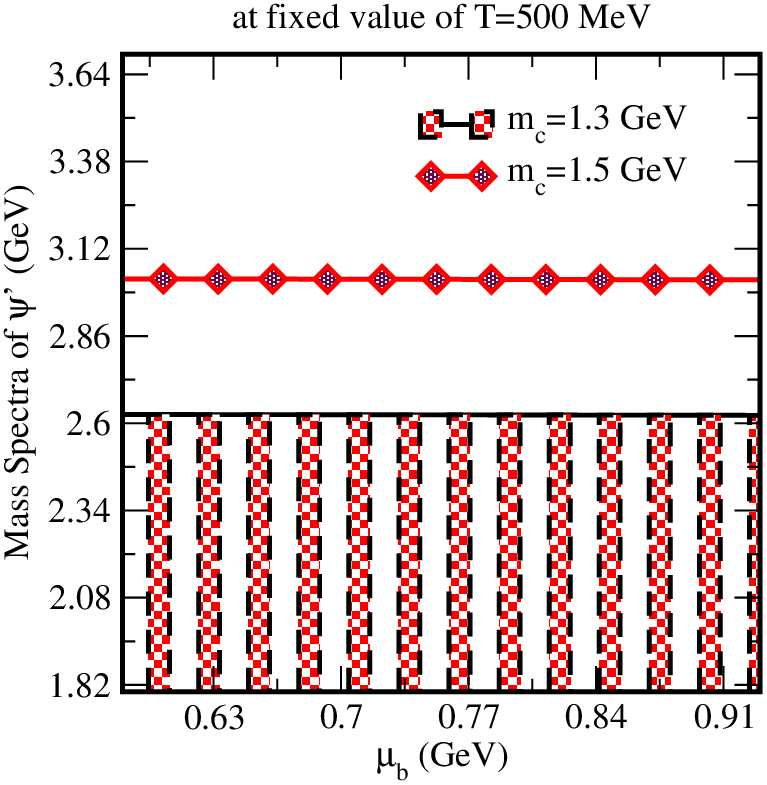}
    \hspace{12mm}
    \includegraphics[height=6cm,width=7cm]{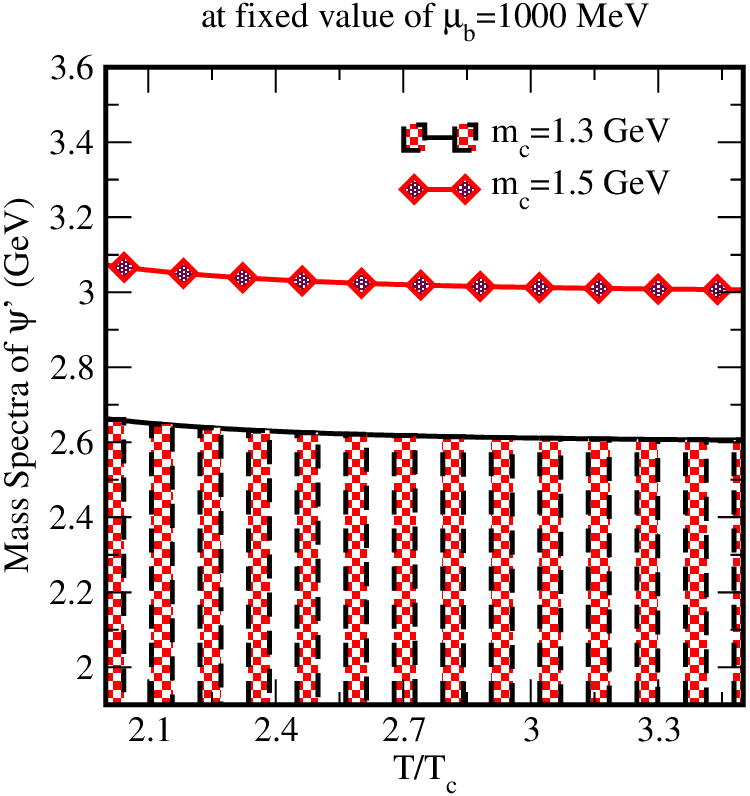}
    \vspace{2cm}
\caption{Dependency of mass of charmonium with $\mu_{b}$ (in left panel) and with $T/T_c$ (in right panel) for excited state of charmonium ($\psi^{\prime}$).}
\label{Figure.9}
 \vspace{3cm}  
\end{figure*}
\begin{figure*}
    \vspace{2mm}   
    \includegraphics[height=6cm,width=7cm]{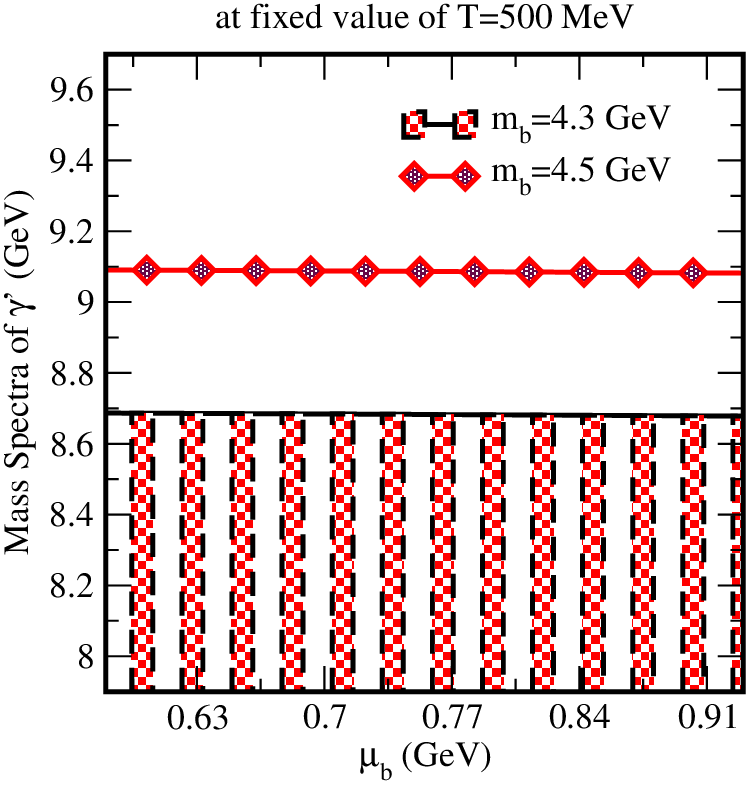}
    \hspace{12mm}
    \includegraphics[height=6cm,width=7cm]{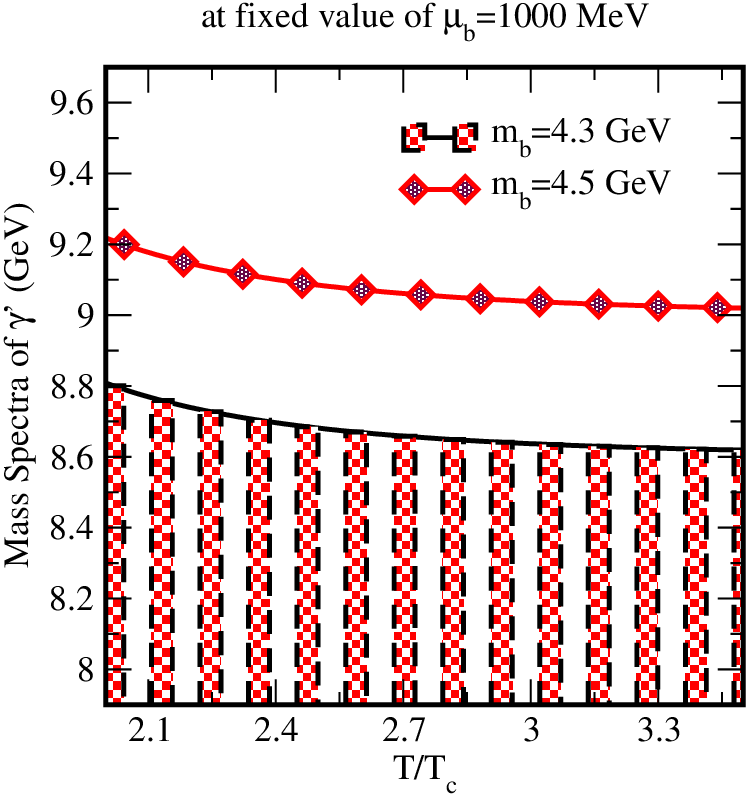}
    \vspace{2cm}
\caption{Dependency of mass of bottomonium with $\mu_{b}$ (in left panel) and with $T/T_c$ (in right panel) for excited state of bottomonium ($\Upsilon^{\prime}$).}.
\label{Figure.10}
 \vspace{3cm}  
\end{figure*}
\begin{table}
\label{table I}
\centering
\caption{The dissociation temperature $(T_D)$ (in GeV) for lower bound state with $T_c$=197 MeV for the different quarkonium states ${J/\psi}$, $\Upsilon$, $\psi^{\prime}$ and $\Upsilon^{\prime}$ has been calculated for the different values of $\mu_{b}$.}
\vspace{0.5mm}
\begin{tabular}{|l|l|l|l|}
\hline
$State$ & $\mu_{b}$=600 MeV & $\mu_{b}$=1000 MeV & $\mu_{b}$=2000 MeV\\
\hline         
$J/\psi$ & 1.5355 & 1.5228  & 1.4720 \\
\hline
$\Upsilon$ & 1.8708 & 1.8654 & 1.7639\\
\hline
$\psi^{\prime}$ & 1.2563 & 1.2561 & 1.2309\\
\hline
$\Upsilon^{\prime}$  & 1.4619 & 1.4593  & 1.4086 \\
\hline
\end{tabular}
\end{table}

\section{Effect of baryonic chemical potential ($\mu_b$) on the mass spectra}
In this section, we have calculated the mass spectra of heavy quarkonium system such as 1S and 2S states of heavy quark and anti-quark for the same value of $N_{f}$=3. For calculating the mass spectra of heavy quarkonia, we have used following relation~\cite{Ibekwe}:
\begin{equation}
\label{eq12}
M=m_{1} + m_{2} + E_{bin}
\end{equation}
Since,
\begin{equation}
\label{eq13}
m_{1} = m_{2} =m_{Q}
\end{equation}
Hence, final expression of mass spectra for the calculation as follow:
\begin{equation}
\label{eq14}
M=2m_{Q} + E_{bin}
\end{equation}
Here, mass spectra is equal to the sum of binding energy ($E_{bin}$) and twice of the quark-masses. Now, we have substituted the values of $E_{bin}$ in the above equation as:
\begin{equation}
\label{eq15}
M=2m_{Q}+\frac{1}{n^2}\frac{m_Q\sigma^2}{m^4_D}
\end{equation}
Where, $m_{Q}$ denotes the masses of 1S and 2S states of quarkonium and $n$ is the principle quantum number.
 
\section{Results and Discussion}
\label{RD}
In this particular work, while considering quasi particle Debye mass at finite temperature and $\mu_{b}$, we have obtained the charmonium and bottomonium binding energy. Figure \ref{Figure.1}, shows that the variation of Cornell potential with distance (r) at a fixed value of $\mu_{b}$ with different values of temperature (left panel) and fixed value of temperature with different values of $\mu_{b}$ (right panel). This potential is not similar to the lattice free energy heavy quark in the deconfined phase, which is well known as coulomb potential \cite{H.Satz}, the Cornell potential solvable by one-dimensional Fourier Transform method in hot QCD medium has similar form that has been used for the study of quarkonium properties, which is considered like color flux tube structure. The variation of QP Debye mass with temperature at different values of $\mu_{b}$ (left panel) and with $\mu_{b}$ at different values of temperature (right panel) respectively has been shown in Figure \ref{Figure.2}. The screening mass at baryon density and temperature was studied by the lattice Taylor expansion method \cite{M.Doring}. In figure \ref{Figure.2}, when we increased the value of $\mu_{b}$, the Debye mass also increased (left panel) and same behavior can be observed for the Debye mass in (right panel) with $\mu_{b}$ for different temperature. The binding energy ($E_{bin}$) of $J/\psi$, $\psi^{\prime}$, $\Upsilon$ and $\Upsilon^{\prime}$ at finite temperature and $\mu_{b}$ with fugacity EoS has been shown in figures \ref{Figure.3}, \ref{Figure.4}, \ref{Figure.5} and \ref{Figure.6}.
\begin{table}
\label{table II}
\centering
\centering
\caption{The dissociation temperature $(T_D)$ (in GeV) for upper bound state with $T_c$=197 MeV for the different quarkonium states ${J/\psi}$ $\Upsilon$, $\psi^{\prime}$ and $\Upsilon^{\prime}$ has been calculated for the different values of $\mu_{b}$.}
\vspace{0.5mm}
\begin{tabular}{|l|l|l|l|}
\hline
$State$ & $\mu_{b}$=600 MeV & $\mu_{b}$=1000 MeV & $\mu_{b}$=2000 MeV\\
\hline         
$J/\psi$ & 1.8274 & 1.8020  & 1.7131 \\
\hline
$\Upsilon$ & 2.2969  &  2.2461 & 2.0930 \\
\hline
$\psi^{\prime}$ & 1.4213 & 1.4086  & 1.3705 \\
\hline
$\Upsilon^{\prime}$  & 1.7258 & 1.7131  & 1.6370 \\
\hline
\end{tabular}
\end{table}
\begin{table}
\label{table III}
\centering
\caption{Comparison of the mass spectra (in GeV) for $J/\psi$ and $\Upsilon$ obtained in the present work with the theoretical and experimental data.}
{\begin{tabular}{|l|l|l|l|}
\hline
$State$ & present work & Exp. mass\cite{tanabashi} & Theoretical mass\cite{Solanki2022}\\
\hline
$J/\psi$ & 3.120  & 3.096 & 3.060\\ 
\hline
$\Upsilon$ & 9.380  & 9.460 & 9.200\\
\hline
\end{tabular}}  
\end{table}
Figures, \ref{Figure.3}, \ref{Figure.4}, \ref{Figure.5} and \ref{Figure.6}, shows the variation of $E_{bin}$ of $J/\psi$, $\Upsilon$, $\psi^{\prime}$, and $\Upsilon^{\prime}$ with $\mu_{b}$ (left panel) and with temperature (right panel) respectively. From these figures it was deduced that, the $E_{bin}$ decreases with $\mu_{b}$ (left panel) as well as temperature (right panel). However, it was clearly observed that baryonic chemical potential has little effect on the binding energy at constant temperature as can be observed from figures \ref{Figure.3}, \ref{Figure.4}, \ref{Figure.5} and \ref{Figure.6}. So, it has becomes an interesting fact about the rate of exponential decay of $E_{bin}$ with increase in the values of $\mu_{b}$. This behavior of $E_{bin}$ can be understood by the strongerness of screening with increase in value of the $\mu_{b}$ and the strength of inter-quark potential is weaker as compared to the case when $\mu_b$=$0$. The $E_{bin}$ at finite value of temperature and $\mu_{b}$ gives information about the dissociation of the quarkonium states (charmonium and bottomonium states). It is known that the $E_{bin}$ is directly proportional to the mass of quarkonia. So, if the value of quarkonia mass increases, this means when we turns towards higher masses i.e. from  $J/\psi$ to $\Upsilon$, we can notice that the binding energy increases. The lower and the upper bound of dissociation temperatures for the quarkonium states $J/\psi$, $\psi^{\prime}$, $\Upsilon$ and $\Upsilon^{\prime}$ is shown in table I and II respectively. It can clearly seen from the table I and II, and with the increasing values of $\mu_{b}$, dissociation temperature $(T_D)$ decreases. The mass spectra of the quarkonia states ($J/\psi$, $\psi^{\prime}$, $\Upsilon$ and $\Upsilon^{\prime}$) with $\mu_{b}$ and temperature is also shown in the left and right panel of the figures \ref{Figure.7}, \ref{Figure.8}, \ref{Figure.9} and \ref{Figure.10} respectively. It was deduced from these figures that if we increases the mass of quarkonium state, mass spectra increases. The values of mass spectra of J/$\psi$ and $\Upsilon$ is shown in table III, and compared with the values of theoretical \cite{Solanki2022} and experimental published data \cite{tanabashi}.\\

\section{Conclusions and outlook}
\label{Con}
We have studied about the quarkonium dissociation pattern in the hot and dense QGP medium, and mapped the quarkonium properties at finite $\mu_{b}$ using medium modified potential. We noticed from the  figures \ref{Figure.3}, \ref{Figure.4}, \ref{Figure.5} and \ref{Figure.6} that the $E_{bin}$ decreases with increasing values of temperature and $\mu_b$. We have also observed that the potential with distance and variation of Debye mass as can be seen from the figures \ref{Figure.1}, \ref{Figure.2} increases with the increasing values of finite temperature and $\mu_{b}$. The behavior of $E_{bin}$ with temperature has also mentioned by the previous studies \cite{prd2016_vin} with anisotropic medium and zero chemical potential in \cite{prd2018_vin}. The dissociation temperature of 1S and 2S state with the $\mu_{b}$, though small variation, and decreases with increase $\mu_{b}$ in the hot QCD medium as shown in table I and II. The mass spectra decreases with both the $\mu_{b}$ and temperature. However, as we increases the mass of the respective quarkonium state, the mass spectra increases as can be seen from the figures \ref{Figure.7}, \ref{Figure.8}, \ref{Figure.9} and \ref{Figure.10}. Such type of studies would contribute to explain the quarkonium properties where the baryon density is very high. Also the dissociation of heavy quarkonia in the presence of baryonic chemical potential is important to find the critical end point (CEP). Facilities like Facility for Anti-proton and Ion Research (FAIR) work on the QGP at such large baryon densities.

\subsection{Data Availability}
The data used to support the findings  of this study is available from the corresponding author upon request .

\subsection{Conflicts of Interest}
The authors declare that they have no conflicts of interest  regarding the publication of this paper.

\section{Acknowledgments}
One of the author V.K.Agotiya acknowledge the Science and Engineering Research Board (SERB) Project No. EEQ/2018/000181 New Delhi for the research support in basic sciences.

\end{document}